# Topological Influence of Sextets on Graphene Oxide Nanostructure

Musiri M Balakrishnarajan[a]*, Gaurav Jhaa[b], and Pattath D Pancharatna[c]*

[a]Chemical Information Sciences Lab, Department of Chemistry, Pondicherry University, Pondicherry-605014, India.
E-mail: mmbkr.che@pondiuni.edu.in

[b]Present Address: Department of Chemical Sciences, Indian Institute of Science Education and Research, Mohali-140306, India.
E-mail: gauravjhaa@iisermohali.ac.in

[c] Department of Chemistry, Amrita Vishwa Vidyapeetham, Amritapuri Campus, Kollam-690 525, Kerala, India.
E-mail: pancharatna@am.amrita.edu

## Abstract

Graphene Oxide (GO) remains a perennial chemical enigma despite its utility in preparing graphene and its functionalization. Epoxides and tertiary alcohols are construed as primary functional groups, but the structural motifs, spectra, and physicochemical properties remain largely inexplicable. Our graph theoretic perturbational analysis of introducing defects (sp$^3$ carbons) on graphene shows the induction of sextets that topologically force 1, 4-quinoidal channels. Edge defects minimize these, whose propagation is halted by vertex "*anti*-defects." Molecular models show that epoxidation creates a diradical centered on two cove-end carbons due to close-lying frontier orbitals. This drives the cascade addition of hydroxides, yielding the ~1:2 ratio of epoxides and hydroxides. This nanostructure accounts for the cation exchange, acidity and anionic nature of GO, and topological conditions that break its σ-framework. The simulated spectra of a 2D sheet with this nanostructure correlate pleasantly with experiments, emphasizing the dominant role of sextet topology on graphene's surface chemistry.

The oxidation of graphite to graphene oxide (GO), arguably the first documented synthesis of a 2D sheet, still eludes chemical comprehension[1,2]. Originally termed as graphitic oxide/acid[3], its 2D monolayers readily exfoliate in various solvents producing a ternary phase of C, O, and H atoms with varying stoichiometry and disorder. Though several models were proposed over time[4], the structural motifs of GO are still shrouded in mystery[2]. The prevalent belief is that epoxides and hydroxides are randomly chemisorbed onto the largely intact hexagonal lattice, leaving some small aromatic islands[5]. Though GO made from different methods[6] widely varies in composition and morphology, $^{13}$C NMR shows commensurability of two peaks, assigned to the tertiary alcohol and epoxide[7] carbons. While oxidative addition on its π network is anticipated, the concomitant presence of these two signals and the unique IR/Raman peaks in all these methods indicate a common structural pattern and local order, potentially driven by unfamiliar site-directing electronic effects, expected of π-electrons. Besides, its preference for hosting strained epoxides[8], origin of acidity disproportionate to the –COOH groups[9], cation exchange capacity[10], etc., remain unexplained. Yet, GO remains the preferred starting material for functionalizing graphene[1,11], including reduced graphene oxide (rGO), due to its low cost and solubility. Ideally, graphene is an infinite polyhex of $sp^2$ carbons whose frontier is dominated by itinerant π-electrons due to negligible *sp* dissonance[12]. It is chemically close to large benzenoids whose π-delocalization is influenced by the peripheral topology and is gauged by Kekulé structures[13] (perfect matchings). Unlike graphene, they show bond alternation to confine π-delocalization within select hexagons[14]. The empirical Clar's sextet theory[15], an off-shoot of resonance theory, effectively captures their chemistry. It is also extended to graphene subsystems[16] in cognizing the preferred topology[17] of their periphery[18]. Here, we encode Clar's approach[19] graph theoretically, to probe the topological impact of π-edge states created internally by oxidizing graphene.

For a hexagonal graph G of a polyhex with n vertices, a Clar Structure is defined as a vertex-spanning sub-graph, having isolated hexagons and edges exclusively, localizing six and two π-electrons, respectively. It is usually drawn on G (C-C σ-framework), with an inscribed circle inside the hexagon and its isolated edges as >C=C< bonds. These Disjoint Resonating Sextets (DRS) structures have every sextet incident to either other sextets (conjugation) or >C=C< bonds. We refer to the number of sextets in its DRS as its ***order*** whose maximum is the quotient of n/6. The DRS with maximum sextets (MDRS) aptly represents its geometry if it is unique; the bond lengths of its isolated edges in >C=C< region (~1.35Å), sextet rings in aromatic region (~1.40Å) and others in $Csp^2$–$Csp^2$ single bonds region (>1.45Å). If multiple MDRS structures exist, their geometry is the superposition of topologically equivalent (degenerate) and non-equivalent (pseudo degenerate) MDRS generated by migrant sextets, represented with arrows[19]. This degeneracy is its ***degree*** (degrees of freedom) whose maximum is the total hexagons in G (linear polyacenes). We refer to the sextets in a DRS as '*in-phase*' if they are directly connected by an edge to show conjugation and '*out-of-phase*' if intervened by a vertex necessitating cross-conjugation. The stability of the polyhex correlates with its MDRS order and, to a lesser extent, by its degree, the latter reducing the frontier gap. The bond alternation due to sextet localization diminishes with the graph size, which narrows the frontier gap. The >C=C< bonds in the MDRS are reactive centers dominating the frontier, and their topology influences the Fermi gap[20].

The ideal infinite graphene is a locally finite, bipartite graph (**G**). Its MDRS is triply degenerate (Degree=3) with maximal order, having sextets exclusively (fully benzenoid[21]), all in phase. Generating these MDRS structures is homeomorphic to the face-coloring of **G** (Extended Data, Figure 1), where each color represents the distinct monophasic MDRS. This uniqueness of its MDRS makes all its hexagons and edges equivalent, hampering sextet localization and band gap. If the hexagon is chosen as the supercell, the valence and conduction bands arising from its π ($e_{1g}$) and π* ($e_{2u}$) crystal MOs are degenerate at Gamma (Extended Data, Figure 2). Unlike 1D polyethyne, which Peierl distorts[22] to one of its two degenerate resonance structures, 2D graphene resists sextet localization, i.e., does not localize into one of its MDRS. The results are equal bond lengths and make it a semi-metal, though graphene shows definite bond-length variations near the periphery.

The sp[3] carbons formed by oxidation will act as defects disrupting π-delocalization and should affect the MDRS, locally inciting bond alternation. Here, we probe the topological impact of saturating a vertex and an edge on **G** using a heuristic[23] that generates sextets width-first around these defects, as its impact will be intense near the defect and fade away radially.

Let a vertex defect be introduced in $\mathcal{G}$ at C1 (Figure 1a). It removes the cyclic delocalization from the three equivalent **A** rings, creating a phenalenyl π-cavity. Its brim generates a [12]-annulene that hosts three cove-type[24] (C2-C3-C6-C3-C2) internal sp[2] edge states (ISES). The three forced single bonds of C1 (C1-C2) prod sextets on the three mutually out-of-phase **B** rings. Since C3 of the adjacent **B** sextets are linked to the common cove-end vertex C6, its DRS must host a >C=C< bond between C6 and C7. Extending further radially requires sextets on **D** rings that force the intervening C8 vertex to mediate their phase difference. This removes the sextet prospects of adjacent **C** & **E** rings. The '*exo*' bonds of the C4 and C5 of the **B** sextet force sextets on the **D** and **G** rings, which recursively extend outward, generating a periodic DRS. This DRS has three localization channels of a 1, 4-quinoidal motif, spreading out in equivalent armchair directions, the C1 defect as the Y-junction. Inserting a single sp[3] defect in G forms an epicentre in π-delocalization, creating a localization wave that cracks open three channels of periodic 1, 4-quinoidal >C=C< bonds acting as fault lines. The resulting DRS (figure 1a) splits the entire lattice into three distinct regions of localized sextets with distinct phases cross-conjugated through >C=C< bonds. This DRS has a reduced order compared to $\mathcal{G}$, but its degree is countably infinite as the >C=C< bonds can be shifted (Extended Data, Figure 3), though its order remains unchanged, making it the MDRS.

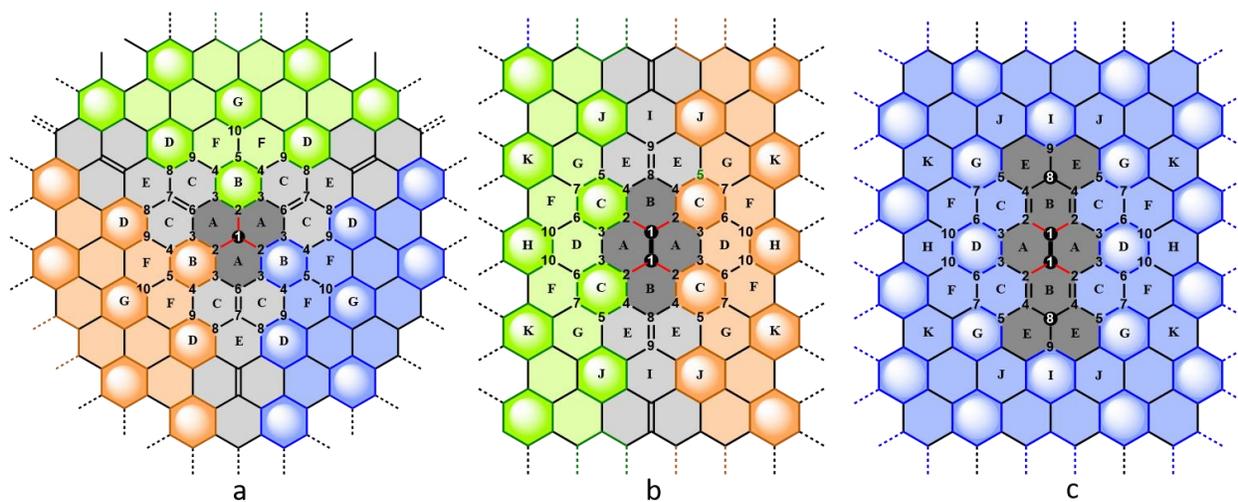

Fig. 1 | The disjoint resonating sextets of the perturbed graphene. The non-degenerate MDRS forced by a, a vertex defect, b, an edge defect, and c, the quenching of the edge defect by two vertex defects. The grey regions show the channels of localized double bonds formed by the internal edges of the cavity (dark grey) surrounding the defect site. The hexagonal rings hosting sextets are shown as shining circles, and their different phases are colored distinctly. Topologically equivalent vertices and rings are marked with the same label.

Figure 1b depicts the sextet localization of the DRS in $\mathcal{G}$ after an edge defect. Saturating the two C1 atoms creates a pyrenyl π-cavity that generates an internal

[14]-annulene ring. It hosts two cove-type ISES (C2-C4-C8-C4-C2) in **B** and two bay-type ISES (C2-C3-C2-C3) in **A**. The four induced C1-C2 single bonds impel sextets on the four **C** rings incident to these $sp^3$ carbons. These sextets, in turn, necessitate single bonds to all exocyclic to **C**, which include the C4-C8 bonds. Having two incident sextets, the cove-end vertex C8 must host the >C=C< bond with C9 to avoid radicalization which restrains **E** and **I** rings from hosting a sextet. This leads to two 1, 4-quinoidal channels, spreading outward in opposite armchair directions. The **C** sextets belong to two distinct phases in the DRS and are out-of-phase across the channels, split by cross-conjugated >C=C< bonds. Moving outward, the in-phase **C** sextets force sextets on the **H** and **K** rings. The localization wave recursively extends with translational periodicity, generating a two-phase DRS. Though this width-first generated DRS is not MDRS (Extended Data, Figure 4), it symmetrically radiates the impact of edge-defect.

The disrupted π-delocalization in 𝒢 by vertex or edge defect destroys the sextet exclusivity of its MDRS by topologically necessitating the destabilizing 1, 4-quinoidal channels, known for inducing radical character[25]. The cove-type ISES is instrumental in forming exo >C=C< bond through its cove-end vertex, initiating the channel formation. Since the vertex defect leads to more localized >C=C< bonds than an edge defect, 1,2-addition is topologically favored. Retrospectively, this also implies that the impact of a defect can be offset by creating additional defects at proper sites. The simple strategy to purge these destabilizing 1, 4-quinoidal channels is to introduce further vertex defects at the cove-end carbons. For the lattice hosting the edge defect (Figure 1b), saturating C8 carbons act as anti-defects halting the formation of extended channels by allowing sextet localization on **I** rings with the missing phase. This phase loses fewer rings in the cavity and becomes MDRS. The resultant terrylene π-cavity (Figure 1c) reinstates the sextet prospects for all hexagonal $sp^2$ rings and restricts the >C=C< bonds of its MDRS (four >C2=C4< bonds) internally within the cavity. Summing up, the impact of an edge defect in 𝒢 by epoxidation can be blocked by two vertex anti-defects, created either by hydroxylation or by capturing two electrons. This explains its anionic nature and the puzzling commensurability of the epoxides with hydroxides observed in all its synthetic reports.

The graph theoretic predictions of the edge defect are tested by analyzing the frontier MOs (FMO) of contrived molecular models. Figure 2 illustrates the correlation of FMOs of the [14]-annulene ring by successive perturbations, computed in their ideal geometries by extended Hückel theory[26].

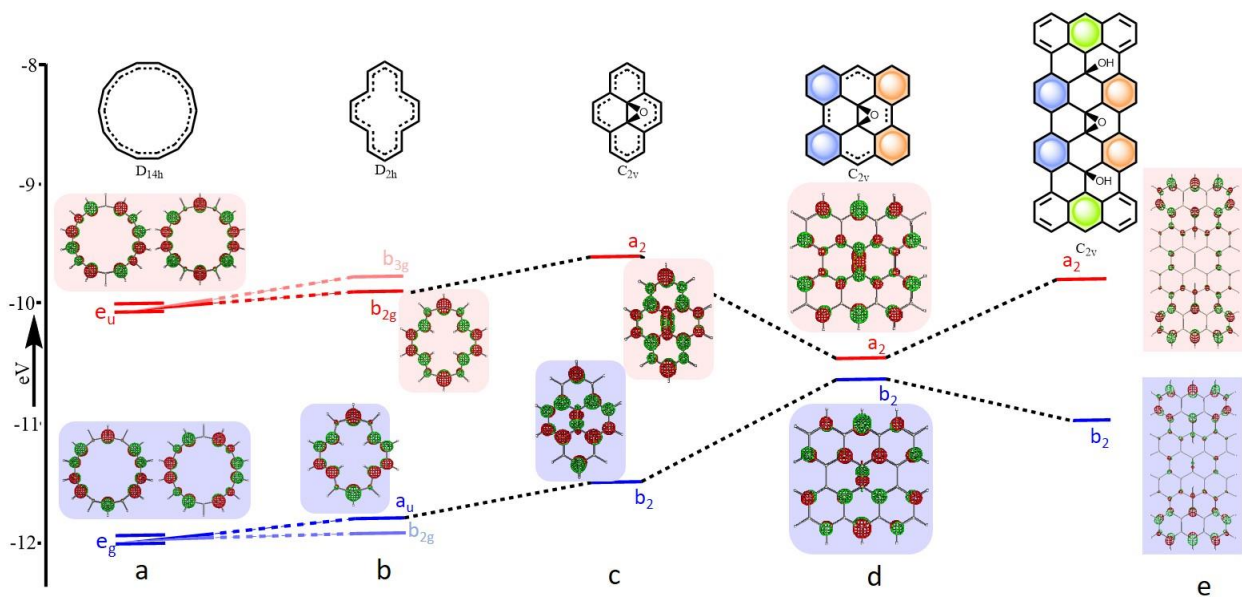

Fig. 2 | Frontier Energy levels of 14-annulene ring subjected to various perturbations a, ideal D$_{14h}$ b, distorted D$_{2h}$ c, perturbation by central epoxide, d, benzannulation to bisanthene epoxide, e, extension to tetranthene with hydroxide groups and epoxide. Occupied and unoccupied levels are colored distinctly.

The degeneracy of HOMO ($e_g$) and LUMO ($e_u$) of the ideal 14-annulene ring (Figure **2a**) is lost with $D_{2h}$ distortion (Figure **2b**). The split-off HOMO ($a_u$) and LUMO ($b_{2g}$) of this pyrenyl π-cavity have large coefficients at the cove-end carbon as expected. Both are slightly destabilized, and the HOMO-LUMO gap is reduced. Introducing a central epoxide (Figure **2c**) to mimic edge defect, destabilizes both these MOs. However, benzannulation to bisanthene epoxide that provides sextet support for the edge defect stabilizes the LUMO ($a_2$) and destabilizes the HOMO ($b_2$). This significantly narrows the gap (Figure **2d**), indicating a reactive diradical[27]. Extending along the 1, 4-quinoidal channel to the tetranthene framework and substituting –OH groups at the cove-end to create anti-defects (Figure **2e**) increases the gap significantly. This reveals the increased kinetic stability of the terrylene π-cavity in the hexagonal network. FMO theory captures the essence, reinforcing the graph theoretic arguments based on resonance theory on edge defect.

The edge defect induced [14]-annulene framework and its isoelectronic analogs were found experimentally to show subdued aromaticity[28]. The dibenzotetraaza [14]-annulene[29] derivatives undergo facile oxidation leading to the supposedly anti-aromatic, 16 π-electron system resembling porphyrin dianion. The cove-end carbons are saturated as tautomers in some dibenzotetraaza[14]-annulene derivatives[30]. Further confirmation of the frontier nature of pyrene epoxide and related systems is obtained from DFT (Density Functional Theory, See Methods) calculations. Geometry optimization of the closed shell singlet of pyrene epoxide breaks the epoxide C-C bond converging to a cyclic ether (Figure 3a), reverting the central carbons to sp$^2$ hybridization. The electrons of the epoxide C-C σ-bond are smoothly shifted to the π-network in the reduced $C_{2v}$ symmetry, rendering the two cove-type ISES into aromatic sextets, restoring the non-degenerate MDRS of pyrene. Its geometry shows negligible bond alternation within the sextets and is conjugated through two >C=C< bonds.

The native pyrene geometry is largely restored, except for the non-planarity inflicted by the strained ether group.

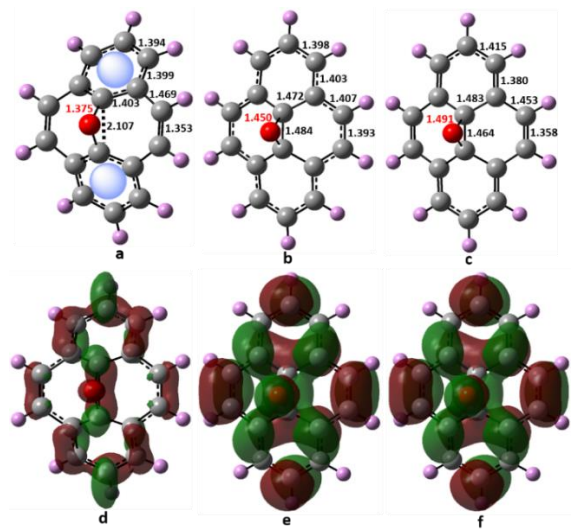

**Fig. 3 | Optimized geometries of pyrene epoxide a,** in its singlet ground state **b,** excited triplet **c,** its singlet dianion, **d,** Evolution of the LUMO of the neutral singlet, **e,** as highest singly occupied orbital of the triplet and **f,** as the HOMO of the dianion.

Though non-planar, the cyclic ether with a pair of sextets is favored over the aromatic [14]-annulene framework with the epoxide. However, its excited state triplet (Figure 3b) and dianion (Figure 3c) stabilize the epoxide form and conserve the σ-backbone, largely retaining its planarity. Remarkably, their minima also conserve the $C_{2v}$ symmetry that forbids bond alternation expected from anti-aromaticity. The saturation of the central C-C bond of pyrene and its photochromic activity is known[31], but its epoxidation-assisted C-C bond breakage is new. The MO that acquires electron(s) is dominated by $p_z$ orbital at the cove-end carbons (Figure 3d-3f). Epoxation induces radicalized Lewis acidity activating these sites, presumably arising from the cove-type ISES topology of the 14-annulene. However, due to the large HOMO-LUMO gap (6.28eV) of the singlet, its triplet is 72.3kcal/mol less stable.

Our systematic inquiry with various benzannulated pyrenes reveals that epoxidation of an internal edge retains the underlying C-C bond whenever it is supported by the surrounding sextet-capable hexagons (C rings in Figure 1b) and the induced cove type ISES (B rings in Figure 1b) are not sextets in their MDRS (Extended Data, Figures 5-18). They turn diradicals if their resulting structure has no perfect matchings[34] (non-kekulean), and their MDRS order is either conserved or increased, revealing the dominating impact of topology. Bisanthene is the smallest framework that sustains the epoxide motif within $C_{2v}$ constraints (figure 4a) but shows a reduced HOMO-LUMO gap (1.52eV) as expected. This leads to wave function instability in its singlet state. Broken symmetry calculations show increased stability for the open shell singlet[32] (23.81 kcal/mol) with substantial diradical character ($y_i$= 0.90)[33], as revealed by occupation numbers of natural orbitals. The Hückel spin density map (Figure 4b) shows the two cove-end carbons having larger α and β spin densities delocalized over the adjacent aromatic rings. Its triplet is slightly (0.86 kcal/mol) more stable than the singlet diradical. The strength of the C-C bond to sustain epoxidation is presumably linked to the increase in the MDRS order from two to four, albeit at the expense of the frontier gap and diradical character. Its dianion is a stable closed-shell singlet with a substantial HOMO-LUMO gap (3.4eV), as expected.

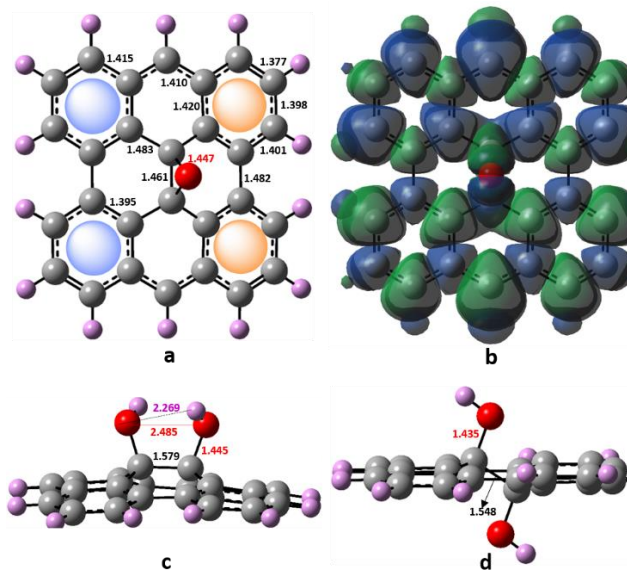

**Fig. 4 |** The bisanthene **a.** geometry of singlet diradical ($C_{2v}$) after epoxidation **b,** its spin density map **c**, geometry of cis diol ($C_2$) and **d**, geometry of trans diol ($C_{2h}$).

Inducing the edge defect in bisanthene by dihydroxylation also yields a diradical showing that the Lewis acidity from the activated LUMO is topological and not specific to epoxidation. The most stable conformation of *syn* glycol (Figure 4c), which hosts -OH groups on the same side, has a triplet ground state (singlet diradical lies 0.01kcal/mol higher) and has steric congestion despite two O-H…O hydrogen bonds which is reflected in its elongated C-C bond. It is less stable by ~15 kcal/mol than its *trans* isomer (Figure 4d). This steric repulsion from eclipsed vicinal –OH groups in *syn* glycol presumably facilitates dehydration leading to epoxide, reducing non-planarity.

Our final model, tetranthene, confirms that the diradical character induced by epoxidation is effectively arrested by hydroxylation of the cove-end carbons (Figure 5). Epoxidation of its central bond yields an open shell singlet diradical (Figure 5a), as expected, which is ~25kcal/mol more stable than its closed shell singlet and marginally less stable (0.77kcal/mol) than the triplet state. Its geometry shows π-localization at the cove end carbons as expected, spreading the spin density along the 1,4 quinoid channel. Saturating the cove-end carbons with the -OH group yields a stable closed-shell singlet (MDRS order six and degree 9) that enables sextet prospects for all hexagonal $sp^2$ carbon rings. The most stable isomer has hydroxides (Extended Data, Figure 19) binding on the same side to epoxide (6.7kcal/mol), with its –OH hydrogens rotated towards the epoxide (Figure 5b). These tertiary alcohols are acidic since their counter ion cyclizes to form epoxides (Figure 5c & 5d) and are stabilized (~1 kcal/mol) despite the ring strain, by reducing non-planarity and delocalizing the charges across the carbon network.

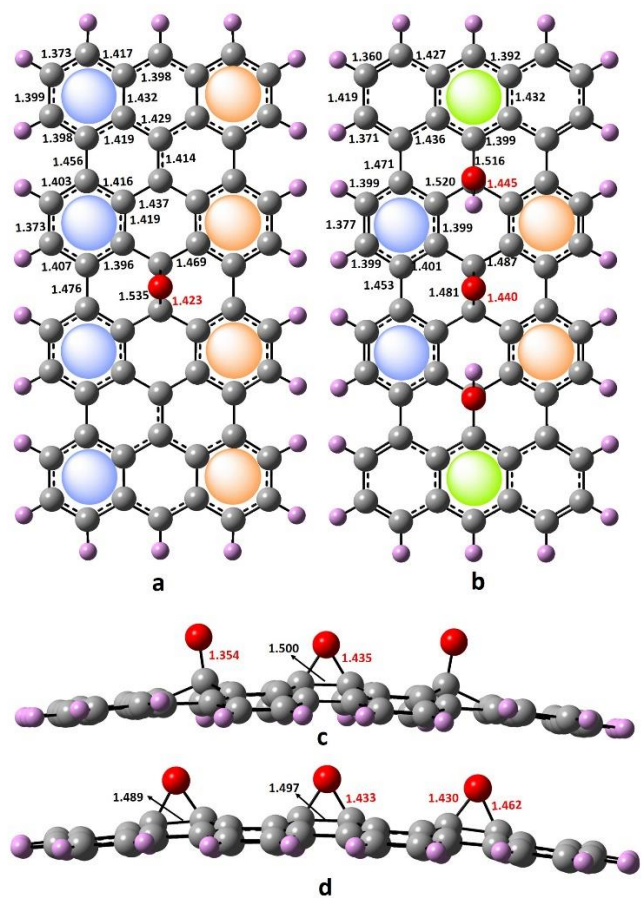

**Fig. 5 | Optimized geometries of tetranthene a,** after epoxidation **b,** after dihydroxylation **c,** its singlet dianion after deprotonation of the O-H group, **d,** its stabilization by forming epoxide bridges.

These results help decode the general oxidation mechanism and the dominant nanostructure of GO. The first step is the reversible *syn* addition of the oxidant through its oxygen bridges on a surface C-C bond; the transition state is stabilized by forming aromatic sextets on hexagons incident to the defective edge. Hydrolysis of this cyclic metal diester intermediate yields cis-diol that undergoes facile dehydration to form epoxide due to steric problems. This syn addition is rapidly followed by hydroxylation at the two cove-end carbons by the oxidant due to its diradical nature, preferably from the same side. Hence, the net reaction involves the cascade addition of epoxide and diol groups through a reactive diradical with frustrated Lewis acidity. This stabilizes its local electronic structure and restores the sextet prospect for all hexagonal $sp^2$ carbon rings. The pyramidalization due to the $sp^3$ defect weakens the π-π interactions with the next graphene layer underneath, facilitating its exfoliation. The unfavorable energetics discourage the ring-opening reactions of these epoxides as the increased pyramidalization of these $sp^3$ carbons reduces the planarity of the residual $sp^2$ framework. Since the diradical formed by epoxidation is quenched by two hydroxides, further oxidation is paced and driven by entropy, presumably leading to non-stoichiometry and disorder. Edges not incident on sextet-capable hexagons have a high reaction barrier as they break the C-C σ-bond. This leads to the strained cyclic ether that will undergo hydrolysis under these conditions forming holes in the lattice. The acidity and anionic nature of GO arise either from the Lewis acidic diradical acquiring electrons from the medium or the dynamic nature[4] of acidic -OH groups as their counter ion is stabilized by forming epoxides.

The idealized building block of GO with translational periodicity with maximum epoxide-diol pair encircled by aromatic sextets is arrived by close packing of the epoxy-diol functionalized terrylene subsystem and has the stoichiometry $C_{16}O_3H_2$ with 25% $sp^3$ carbons. DFT calculations show epoxides and diol pairs prefer being on the same side due to reduced puckering of the carbon lattice, as in Figure 5c, while adjacent epoxy-diol pairs alternate on either side to avoid steric problems (Extended Data, Figure 20-21).

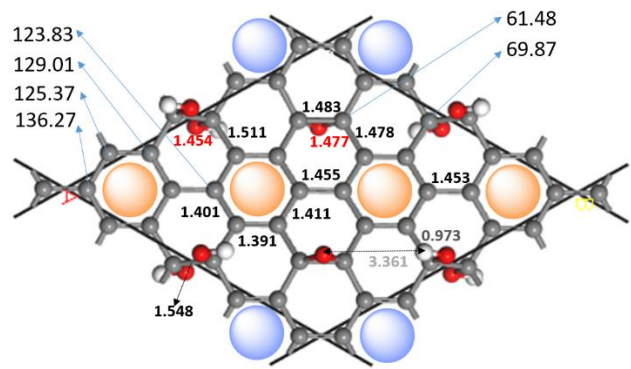

Fig. 6 | Optimized geometry of the idealized GO unit cell with maximal epoxide-diol pairs separated by aromatic rings. The C-C and C-O bond lengths are shown in distinct colors. The Chemical shifts of symmetrically distinct carbons are indicated by (blue) arrows.

The most stable unit cell of GO has *poly (p-phenylene)* chains interleaved by parallel epoxide-diol pairs alternating on either side of the sheet in a centered rectangular lattice of *c2mm* 2D space symmetry (Figure 6). The optimized geometry of its primitive cell (Z=2, a=9.894 Å, α = 120.18) has an essentially planar $sp^2$ framework (deviation < 0.2Å), while the $sp^3$ carbons deviate from the average plane by ~0.5Å, with their bond-lengths within the expected range.

The computed isotropic $^{13}C$ chemical shifts calibrated using benzene as the reference show epoxide carbons at 62 ppm and hydroxyl carbons at 70 ppm, while the aromatic carbons appear with an average chemical shift of 132 ppm, in exceptional agreement with the experiments[35]. It also confirms the earlier predictions of the aromatic nature of all $sp^2$ carbons, the absence of isolated >C=C< bonds, and large graphenic regions[36]. Besides, the current skeleton is consistent with the observed cross-peaks between $sp^3$-$sp^2$ carbons experimentally reported in 2D NMR and the absence of vinyl $sp^2$ carbons between epoxy-diol pairs indicated by their proximity[37].

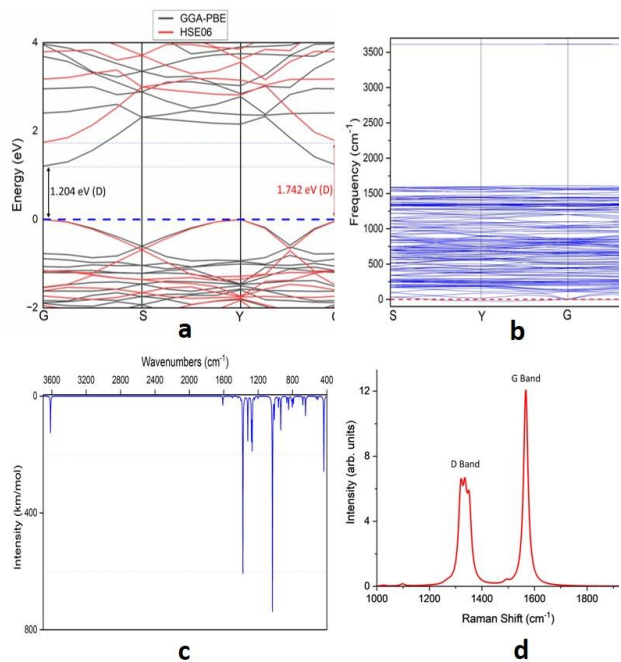

Fig. 7 | The Model GO sheet (a) Electronic band structure, (b) Phonon Dispersion, (c) IR, and (d) Raman spectra from DFT calculations.

The structure (Figure 7a) has a direct band gap at Gamma with the frontier bands primarily dominated by poly (p-phenylene) chains (Extended Data, Figure 20), as expected[20]. Phonon dispersion (Figure 7b) shows no imaginary modes at Gamma, indicating its dynamic stability[38]. The computed vibrational spectra show the most intense infra-red peaks at 1033 cm$^{-1}$ ($B_{2u}$) and 1376 cm$^{-1}$ ($B_{3u}$), (Figure 7c) and Raman $A_g$ peaks appear at 1319 cm$^{-1}$, 1335 cm$^{-1}$, 1351cm$^{-1}$ expected of D-bands, and 1567 cm$^{-1}$ for G band (Figure 7d) in incredible agreement with the experiments[39,40]. The surprisingly excellent correlation of multiple experimental spectral data points to the prevalence of epoxide-diol functionalized terrylene cavity as the primary structural motif for the nanostructure of GO.

Oxidation beyond this stoichiometry presumably requires disrupting the sextets. Further defects will break the conjugation in polyphenylene chains. Though the bonds linking phenyl groups have favorable steric accessibility, an edge defect will break the underlying σ-bond to acquire sextet support, as indicated by the pyrene model (Figure 2). The degree of oxidation of this ideal periodic structure is less compared to most experimental data, presumably due to the persistence of adsorbed water and frequent breakdown of the hexagonal C-C σ-framework by over-oxidation.

## Conclusion

By analyzing the impact of introducing $sp^3$ carbon defects on the π-delocalization of graphene, this probe establishes the preference for epoxidation and topological requirement of tandem dihydroxylation, on the cove-end vertices of the ISES. This functionalized terrylene motif is the primary building block for the nanostructure of GO, as it correlates well with various experimental spectra despite its superficially assumed ideal periodicity. Graph theoretic analysis based on the VB/resonance theory of Clar, frontier MO theory, DFT calculations on model molecules/2D sheets, and the computed spectra, all reinforce this conclusion. This model ably explains all the observed anomalies, i.e., the facile formation of robust epoxide despite the ring strain, the observed commensurability of epoxide and hydroxides, the perplexing acidity of these tertiary alcohols, and its cation-exchange nature that are common to GO across different preparation methods. It also accounts for the formation of larger holes due to over-oxidation and the conditions that break the hexagonal carbon framework. The insights drawn here on the nanostructure and mechanistic formation of GO unveil the importance of sextets and their topology.

These results will help engineer finer control over the reactions of graphene surface that aids rational synthesis of functionalized graphene sheets.

**Online content**
Any methods, additional references, Nature Portfolio reporting summaries, source data, extended data, supplementary information, acknowledgements, peer review information; details of author contributions and competing interests; and statements of data and code availability are available at https://doi.org/xxx

*Phys. Chem. C.*, **119**, 10123-10129 (2015).

**Methods:** The electronic structure and geometry of molecular models were obtained from calculations from Density functional theory employing the MO6-2X exchange-correlation functional with the 6-311+G (d,p) basis set within the Gaussian 16 software.[41] This method is used for the geometry optimization and characterization of the nature of stationary points of singlets, triplets and dianions on their potential energy surface. The obtained the wave functions were tested for their stability, and if instability exists, the broken symmetry uMO62x method was used to obtain accurate energies of the diradical.

Periodic nanosheet calculations were conducted using the Cambridge Ab initio Serial Total Energy Package (CASTEP)[42] program, applying density functional theory. The Perdew-Burke-Ernzerhof (PBE)[43] functional within the non-local corrected generalized gradient approximation (GGA)[44] was utilized to acquire optimized geometry, energies, and band structure data. Grimme's dispersion correction[45] was also incorporated to account for potential weak interactions involving oxygen lone pairs and π-electrons. The methodology involved the implementation of the BFGS (Broyden–Fletcher–Goldfarb–Shanno) algorithm,[46] ultrasoft pseudopotentials, and a plane-wave basis set with a kinetic energy cutoff of 500 eV in a Monkhorst-Pack k-mesh,[47] maintaining a separation of 0.04 1/Å along the two lattice directions for geometry optimization, band structure analysis, and density of states calculations. The adjacent nanosheets were kept at a minimum distance of 20 Å to prevent possible long-range interactions. The well-converged geometries for the nanosheets were obtained by simultaneous optimization of individual atom positions and lattice parameters while adhering to the following convergence criteria: SCF tolerance ($5.0 \times 10^{-6}$ eV/atom), maximum stress (0.02 GPa), maximum displacement ($5.0 \times 10^{-4}$ Å), and forces (0.01 eV/Å). We also used the HSE06 (Heyd-Scuseria-Ernzerhof 2006) hybrid functional for refined band structure calculations. The same dispersion-corrected GGA formulation (as used in geometry optimization) was employed for phonon dispersion analysis, Raman and IR spectrum, utilizing norm-conserving pseudopotentials[48] in conjunction with the linear response method[49] and Koelling-Harmon relativistic treatment.[50] The plane-wave cutoff was increased to 830 eV while maintaining a Monkhorst–Pack k-mesh separation of 0.04 Å for improved accuracy. We used a standard Ar laser (514.5 nm) as incident light and a Lorentzain broadening width of 20.0 cm-1 for the Raman spectrum.

NMR shielding tensors for the periodic system were computed with the gauge including the projector-augmented plane-wave (GIPAW) method[51,52] implemented by Pickard and Mauri[51] and extended to OTFG ultrasoft pseudopotentials[53] as well as the revised PBE (RPBE) nonhybrid gradient density functionals[54-57].

As a commonly used method, chemical shifts were calibrated with benzene as an internal reference,[58,59] which generally leads to a quantitative agreement between theory and experiment. The isotropic nuclear shielding of the benzene molecule was calculated at 44.74 ppm, which agrees well with previous studies.[58-60]

We have applied the PBE and the RPBE functionals and used TMS ($\delta_{TMS}^{TMS}$) and benzene ($\delta_{Benzene/TMS}$) as the NMR references with

$$\delta_{GO}^{TMS} = \sigma(\text{Benzene})^{iso} - \sigma(\text{GO}) + \delta_{GO}^{benzene}$$

Where the $\delta_{GO}^{TMS}$ is the chemical shift of graphene oxide (benzene as internal reference) σ(Benzene)$^{iso}$ is the isotropic nuclear shielding of the benzene, σ (GO) is the chemical shift of graphene oxide lattice, and $\delta_{GO}^{benzene}$ is the experimental chemical shift for benzene (126.9 ppm)[58-60].

**Data availability**

Raw data for key computational results are presented in Extended Data Figs. 1-18. Further data can be requested from the corresponding author.

**Acknowledgments:** MMB acknowledges the initial support from DST, India (Grant No. SR/S1/PC-10/2007), PDP for the Women Scientist Scheme (WOS-A) (Grant no. CS-130/ 2016), and G.J. acknowledges the CSIR, India, for the NET Fellowship.

**Author Contributions:** MMB conceived this project through graph theoretic analysis and the selection of computational models. GJ performed DFT calculations and PDP analysed the electronic structure. MMB prepared the manuscript using input from all co-authors.

**Competing interests:** The authors declare no competing interests.